\documentclass[oldversion]{aa}
\usepackage{amsmath,graphicx,amssymb}

\usepackage{natbib}
\bibpunct{(}{)}{,}{a}{}{,}
\defcitealias{nltei}{KKI}

\newcommand{\mdot}{\ensuremath{\dot M}}
\newcommand{\Teff}{\mbox{$T_\mathrm{eff}$}}
\newcommand{\zav}[1]{\left(#1\right)}
\newcommand{\ms}{\ensuremath{\text{M}_{\odot}}}
\newcommand{\kms}{\ensuremath{\mathrm{km}\,\mathrm{s}^{-1}}}
\newcommand{\msr}{\ensuremath{\ms/\text{year}}}

\newcommand\de{\text{d}}
\newcommand{\lx}{\ensuremath{L_\text{X}}}

\allowdisplaybreaks

\begin{document}

\title{X-ray emission from hydrodynamical simulations in non-LTE wind models}
\titlerunning{X-ray emission from hydrodynamical simulations in non-LTE wind
models}

\author{J.  Krti\v{c}ka\inst{1} \and A. Feldmeier\inst{2} \and
L. M. Oskinova\inst{2} \and J. Kub\'at\inst{3} \and W.-R. Hamann\inst{2}}
\authorrunning{J. Krti\v{c}ka et al.}

\institute{\'Ustav teoretick\'e fyziky a astrofyziky, Masarykova univerzita,
            Kotl\'a\v rsk\'a 2, CZ-611 37 Brno, Czech Republic,
	    \email{krticka@physics.muni.cz}
           \and
           Institut f\"ur Physik und Astronomie, Universit\"at Potsdam,
           Karl-Liebknecht-Stra{\ss}e 24/25, 14476 Potsdam-Golm, Germany
           \and
           Astronomick\'y \'ustav, Akademie v\v{e}d \v{C}esk\'e
           republiky, CZ-251 65 Ond\v{r}ejov, Czech Republic}

\date{Received: }

\abstract{Hot stars are sources of X-ray emission originating in their
 winds. Although hydrodynamical simulations that are able to predict
 this X-ray emission are available, the inclusion of X-rays 
 in
 stationary wind models is usually based on simplifying approximations.
 To improve this, we use results from time-dependent hydrodynamical
 simulations of the line-driven wind instability
 (seeded by the base perturbation) to derive the
 analytical approximation of X-ray emission in the stellar wind. We use
 this approximation in our non-LTE wind models and find that an improved
 inclusion of X-rays leads to a better agreement between model
 ionization fractions and those derived from observations. Furthermore,
 the slope of the $L_\text{X}-L$ relation is in better agreement with
 observations, however the X-ray luminosity is underestimated by a
 factor of three. We propose a possible solution for this discrepancy.

  }

\maketitle

\section{Introduction}

Hot-star winds have been traditionally modelled assuming a spherically
symmetric, stationary outflow. These assumptions provide a convenient
basis for studying different phenomena that may influence the wind
structure. The line profiles and wind parameters predicted in this way
are in good agreement with those found from observations \citep
[e.g.,][]{pahole,vikolamet,nlteii}.

However, a stationary approach to hot-star wind modelling was for a long
time known to be not fully adequate \citep{lusol}. Hydrodynamical
simulations show the growth of strong shocks in the wind due to the
so-called line-driven instability inherent to radiative driving
\citep{ocr,felpulpal}. On the other hand, from the observational point
of view the inhomogeneities do not imprint a clear signature in hot-star
wind spectra in ultraviolet and visible spectral regions, and only a
detailed spectral analysis 
reveals
its possible presence
\citep{bourak,martclump,pulchuch}. Consequently, the influence of
inhomogeneities (or clumping) on wind spectra has long been neglected.
At present it is not clear whether the neglect of wind clumps in the
formation of wind line profiles and infrared continua leads to an
overestimate of mass-loss rates derived from observations
\citep[see][for a discussion of this problem]{pulamko}.

The only clear observable signature of intrinsic wind non-stationarity is the
existence of X-ray emission. This emission is directly observable for nearby
stars \citep[e.g.,][]{rosatvel,igorkar} and its existence can be inferred from
the influence on the ultraviolet spectrum for those stars for which a direct
observation of their X-ray emission is not available \citep[e.g.,][]{lucka}.

The line-driven instability is not the only possible production
mechanism of X-rays in hot star winds. Because the winds are highly
supersonic, any mechanism which causes wind stream collisions may also
lead to X-ray production. Consequently, in hot star binaries the X-rays
may originate due to the collisions of winds from individual stars
\citep[e.g.,][]{usaci,kapitan,skoda}. The collisions of individual wind
streams 
channeled
by the magnetic field may also cause X-ray emission
\citep{bamo,udo}. The former mechanism is important in binaries and the
latter one in stars with 
a
sufficiently strong magnetic 
field.
In this
paper we concentrate on a more general mechanism, operating in all
radiatively driven stellar winds, i.e., 
%
the line-driven instability.

A central observational finding related to the X-ray emission is the
dependence of the X-ray luminosity $L_\text{X}$ on the total luminosity
$L$ via the approximate relation $L_\text{X}\sim10^{-7}L$ \citep
[e.g.,][] {chleba}. This relationship is still not explained by wind
theory. Assuming a constant X-ray filling factor, \citet{oskal} showed
that for stars with optically thick winds the observed X-ray luminosity
scales with the mass-loss rate as $\lx\sim\mdot$, and for stars with
optically thin winds as $\lx\sim\mdot^2$. As the predicted wind
mass-loss rate scales with the stellar luminosity as $\mdot\sim
L^{1/\alpha'}$ \citep{kupul}, where $\alpha'\approx 0.6$ for luminuous O
stars\footnote{Note that the situation may be different in late O and
B supergiants, for which the value $\alpha'\approx 0.8-1$ might be a
more appropriate one \citep{banag}.}
\citep[e.g.,][]{vikola,nlteiii,pulvina}, the predicted slope of the
$\lx-L$ relation is steeper than the observed one. This difference may
originate in the radial dependence of 
the
X-ray filling factor \citep{oskal}
or may be caused by the dependence of the cooling length on the wind
density \citep{nlteiii}. Note also that macroclumping
\citep{por1,osfeha,owosha,xlida} may lead to a decrease in the effective
opacity in the X-ray region and affect the X-ray luminosity.

Although a real hot-star wind should be far from 
%
stationary and
spherically symmetric (on small scales), the success of wind modelling
applying these approximations motivates us to incorporate the
inhomogeneities found in time-dependent hydrodynamical simulations into
a spherically symmetric, stationary wind model. These models can serve
as an efficient tool to study hot-star winds until more elaborate
hydrodynamical simulations that consistently take into account the
non-LTE effects become available.

There 
have been
earlier attempts to include X-ray emission 
in
non-LTE wind
models. They were either based on simplified analytical models
\citep{felkupa,martclump}, or the X-ray emission was included using
ad-hoc free parameters \citep[aka the "filling factor", e.g.,][]
{macown,martclump,nlteiii} describing the hot wind part. Here we use the
results of available hydrodynamical simulations of \citet{felpulpal} to
directly describe the X-ray emission in a compact form and include it in
our non-LTE wind models.

\section{Hydrodynamical simulations}

The hydrodynamical simulations we employ here to estimate the X-ray
emission from O\,stars were performed using the {\it smooth source
function} method \citep[SSF,][]{Owocki1,Owocki2}. This
corresponds roughly to a formal integral of the radiation force using
a pre-specified line source function, here the optically thin source
function, $S\sim r^{-2}$. The line-driven instability that is
responsible for the formation of X-ray emitting shocks in the wind is
captured by a careful integration of the line-optical depth with a
resolution of three 
observer-frame
frequency points per line Doppler
width and accounting for non-local couplings in the non-monotonic
velocity law of the wind \citep[see][for details]{ocr}. The
angle integration in the radiative flux is limited to one single ray
which, however, is not the radial ray but hits the star at $\approx
0.7R_\ast$. This accounts with sufficient accuracy for the finite-disk
correction factor \citep{Pauldrach} and avoids the
critical-point degeneracy of the point-star CAK model \citep{Poe}.

To avoid any artificial overestimate of the X-ray production in the wind, special
care was taken in the numerical calculation of the line force on the staggered
spatial mesh in order to correctly reproduce the line-drag effect \citep{Lucy},
which is responsible for a partial reduction of the 
instability,
especially in
layers close to the star \citep{Owocki4}. The pure hydrodynamics part of the
code is a standard van Leer solver following standard prescriptions for such a
scheme: staggered mesh; operator splitting of advection and source terms;
advection terms in 
a
conservative form using van Leer's (\citeyear{vanLeer})
monotonic derivative as an optimised compromise between stability and accuracy;
Richtmyer artificial viscosity; and non-reflecting boundary conditions
\citep{Hedstrom,Thompson1,Thompson2}. The numerical collapse of post-shock
radiative cooling zones caused by the strong oscillatory thermal instability
\citep{Langer} is prevented by artificially modifying the radiative cooling
function below a certain temperature at which X-ray emissivity is still small.
Further details can be found in \citet{Feldmeier}.

As seed perturbations for unstable growth we introduce a turbulent
variation of the velocity at the wind base, at a level of roughly one
third of the sound speed. This perturbation is obtained from a simple
quadrature of the Langevin equation \citep[see][]{Risken} with a coherence
time in the friction term of 5000 seconds, which is well below but not
too far off the acoustic cutoff period of the star at which, according
to 
the
theorem by Poincare \citep[see][]{Lamb}, acoustic perturbations of
the photosphere should accumulate. The power spectrum $E(k)$ of this
Langevin turbulence has a spectral index of $-2$, which is close to
the Kolmogorov index of $-5/3$ for the universal, inertia subrange of
turbulence.

The variety of dynamical structures in the instability-induced,
line-driven wind turbulence (which grows out of but is not identical
to the turbulence applied in the photosphere via boundary conditions)
is still largely unexplored, but seems to be similarly intricate
to
that found in other turbulent flows in ordinary fluids and gases or in
astrophysical MHD settings. There are indications of the presence of a
quasi-continuous 
hierarchy
of density and velocity structures in the
wind, similar to that found in supersonic Burgers turbulence with
shock cannibalism \citep{Burgers}, and there are also indications of a
separation of dense structures into two distinct families, which we
address under the names ``shells'' and ``clouds.''  The shells are
formed close to the star, in a first stage of unstable growth. They do
not collect all the wind material but rather one half of it, since the
negative-velocity perturbations remain unaffected by the line-driven
instability \citep{MacGregor}. Further out in the wind, in a
second stage of unstable growth, clouds are ``ablated'' from the outer
rim of the remaining mass reservoir at CAK densities, and are
accelerated by the stellar radiation field through the 
emptied
regions, eventually colliding with the next-outer shell, and producing
observable X-ray flashes in this collision \citep{felpulpal}.

Without photospheric Langevin turbulence, clouds do not occur and the
mass reservoir is continuously fed into the next outer shell via a
thin stream of fast gas that 
by far cannot 
\citep[see][]{Hillier}
account for the observed X-ray emission from O stars. To
test for the 
%
relevance of the specific form of photospheric
turbulence, we also calculated models with a ``stochastic''
photospheric sound wave as seed perturbation for the instability,
i.e.~a sound wave with stochastic variations in period, amplitude, and
coherence time, which resulted in largely the same results.

\section{X-ray emission from hydrodynamical simulations}

To incorporate the results of hydrodynamical simulations of
$\zeta$~Ori~A wind in non-LTE wind code in a manageable way, we
approximate the emission from hydrodynamical simulations as a polynomial
function. This could be done in two ways.  The first way is to calculate
the X-ray emission based on a local values of hydrodynamical variables,
and then determine a function that best approximates its radius and
frequency dependence (see Sect.~\ref{kapspec}). The second, somewhat
simpler way, is to find a polynomial that fits the temperature structure
of the simulation, and then use this polynomial to calculate the
emission (see Sect.~\ref{kaptem}).

\subsection{Spectrum fitting}
\label{kapspec}

\begin{figure*}
\centering
\includegraphics[width=0.49\hsize]{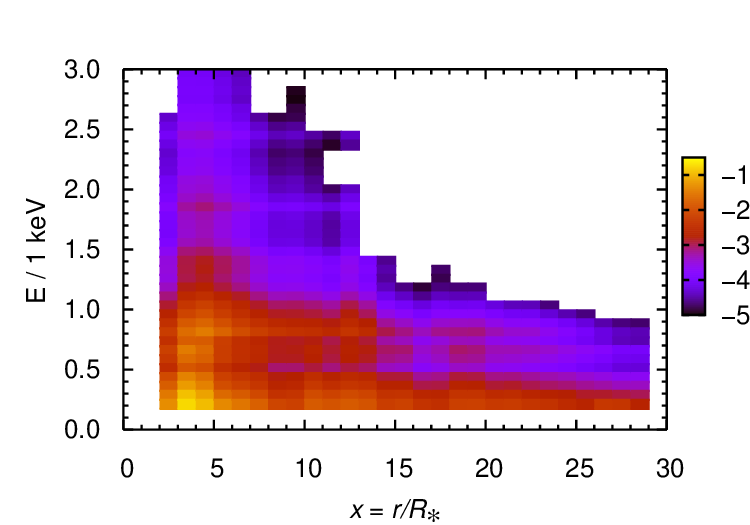}
\includegraphics[width=0.49\hsize]{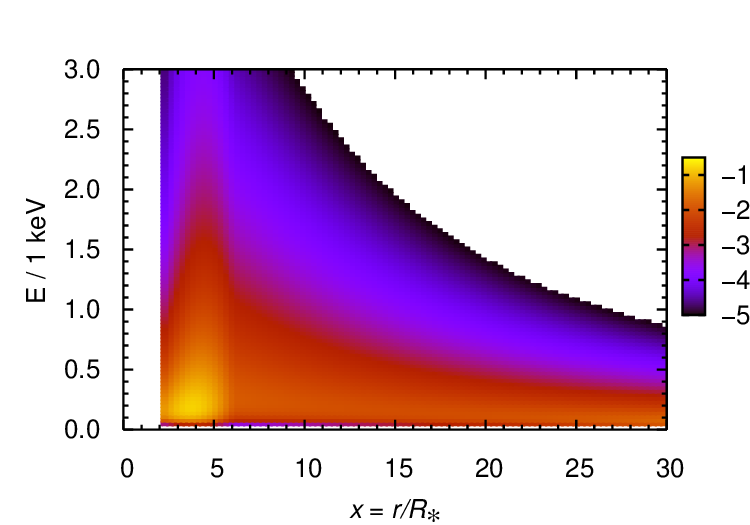}
\caption{{\em Left}: 
  The frequency distribution of X-rays expressed in terms of $\log
  (\ell_{ij}/10^{-23}\,\text{erg}\,\text{s}^{-1}\,\text{cm}^{-3}\,
  \text{keV}^{-1})$ as a function of radius and frequency (see
  Eq.~\eqref{ellir}).  {\em Right}: Fit to the X-ray emissivity
  expressed as $\log \ell$ (Eq.~\eqref{frefit}).}
\label{hyd2fil}
\end{figure*}

First, the wind structure from hydrodynamical simulations is used to
calculate the X-ray emissivity as a function of radius and frequency.
We divide the wind model into 29 equally spaced radial bins and
calculate the total wind X-ray emission in each radial bin at the
timestep $k$ as an integral
\begin{equation}
l_{ijk}=\int_{r_{i}}^{r_{i+1}}
r^2 n_{\text{e}}(r)n_\text{H}(r)\Lambda(\nu_j,T(r)) \,\de r,
\end{equation}
where the radius $r_{i}$ denotes the radial boundary of the $i$-th
bin, $n_{\text{e}}(r)$, $n_\text{H}(r)$ are the number densities of
free electrons and hydrogen, and the X-ray emissivity
$\Lambda(\nu_j,T(r))$ is calculated using the Raymond-Smith X-ray
spectral code (Raymond \& Smith \citeyear{rs}, Raymond \citeyear{ray})
for uniformly spaced frequencies $\nu_j$ with the temperature $T(r)$.
The variables $n_{\text{e}}(r)$, $n_\text{H}(r)$, and $T(r)$ were
taken from the simulation of $\zeta$~Ori~A wind \citep{felpulpal}.
We calculate the quantities
\begin{equation}
\label{ellir}
\ell_{ij}= \frac{1}{\Delta\nu}\frac{\overline{l}_{ij}}{m_{i}},
\end{equation}
where 
\begin{equation}
m_{i}= \int_{r_{i}}^{r_{i+1}}
r^2\tilde n_{\text{e}}(r)\tilde n_\text{H}(r)\,\de r,
\end{equation}
$\overline{l}_{ij}$ is the time average of $l_{ijk}$, and $\Delta\nu$ is
the frequency spacing. 
%
$\tilde n_{\text{e}}(r)$ and $\tilde
n_\text{H}(r)$ are the electron and hydrogen number densities from a
stationary wind model corresponding to hydrodynamical simulations. This
stationary model is calculated by adopting the mean mass-loss rate from
the hydrodynamical simulations \citep{felpulpal} and using a standard
beta velocity law.

To provide a simpler description of the X-ray 
emission,
we fit the
values of $\ell_{ij}/10^{-23}\,\text{erg}\,\text{s}^{-1}\,
\text{cm}^{-3} \,\text{keV}^{-1}$ by (see Fig.~\ref{hyd2fil})
\begin{equation}
\label{frefit}
\log \ell(x,\gamma)=a(x)+b(x)\gamma+c(x)\gamma^2+d(x)\gamma^3,
\end{equation}
where $\gamma=\log(E/1\,\text{keV})$ is the logarithm of energy $E$
expressed in 
%
keV, and $x=r/R_*$. The radial variation of
the polynomial coefficients is given by the expressions
\begin{equation}
\label{zosen}
\begin{split}
a(x)&=-6.087+1.836x_6-0.211x_6^2-0.112
\zav{x-x_6},\\
b(x)&=-4.124+0.232x_6-0.132
\zav{x-x_6},\\
c(x)&=0.651-0.456x_6,\\
d(x)&=2.457-0.369x_6,\\
x_6&=\left\{\begin{array}{c}
               x,\qquad x<6,\\
	       6,\qquad x\geq6.\\
	       \end{array}\right.\\
\end{split}
\end{equation}
Note that $\ell=0$ for $x<2$. The X-ray emissivity 
in the non-LTE model can be then approximated by
\begin{equation}
\label{etax}
\eta_\text{X}(r,\gamma)=\frac{f^*}{4\pi} n_\text{e}n_\text{H}
\ell(r/R_*,\gamma)\, 10^{-23}\, \text{erg}\,\text{s}^{-1}\,
\text{cm}^{-3}\,\text{keV}^{-1},
\end{equation}
where $n_\text{e}$, $n_\text{H}$ are the number densities of free electrons and
hydrogen in the non-LTE model. $f^*$ is the scaling factor (see below,
Eq.~\eqref{kolin}), which allows
one
to use Eq.~\eqref{etax} also for
stars other
than $\zeta$~Ori~A. In Eq.~\eqref{etax}
we introduced the factor $1/4\pi$ because the emissivity
$\Lambda(\nu,T)$ is assumed to be that 
in
all spatial directions. 

\subsection{Temperature distribution fitting}
\label{kaptem}

Instead of fitting the X-ray emissivity from simulations directly, it
is also possible to fit the temperature distribution of X-ray emitting
gas. This approach does not explicitly depend on a particular
functional form of the X-ray emissivity $\Lambda(\nu,T)$, but is also not
completely independent of it, as the distribution of temperatures in
the numerical simulations depends on the adopted form of the cooling
function.

As the amount of emitted X-rays per unit volume of gas depends on the
square of the density $n^2$, it is not realistic to fit just the
distribution of temperatures. To account for the correlation of the
distribution of $n^2$ with temperature, we evaluate for each timestep
$k$ the quantity similar to the differential emission measure
\citep[e.g.,][]{radon}
\begin{equation}
\text{DEM}_{ijk}=
\int_{r_{i}}^{r_{i+1}} r^2 n_{\text{e}}(r)n_\text{H}(r)\delta_j(T(r))\,\de r,
\end{equation}
where $\delta_j(T(r))$ is equal to 1 for $T_j-\Delta T_j/2 < T(r) <
T_j+\Delta T_j/2$ and 0 otherwise. Here we introduced the
logarithmically spaced grid of temperatures $T_j$ with grid size
$\Delta T_j$. The normalised DEM is introduced as
\begin{equation}
\mathcal{D}_{ij}=\frac{1}{\Delta T_j}\frac{\overline{\text{DEM}}_{ij}}{m_{i}},
\end{equation}
where $\overline{\text{DEM}}_{ij}$ is the time average of
$\text{DEM}_{ijk}$.

\begin{figure*}
\centering
\includegraphics[width=0.49\hsize]{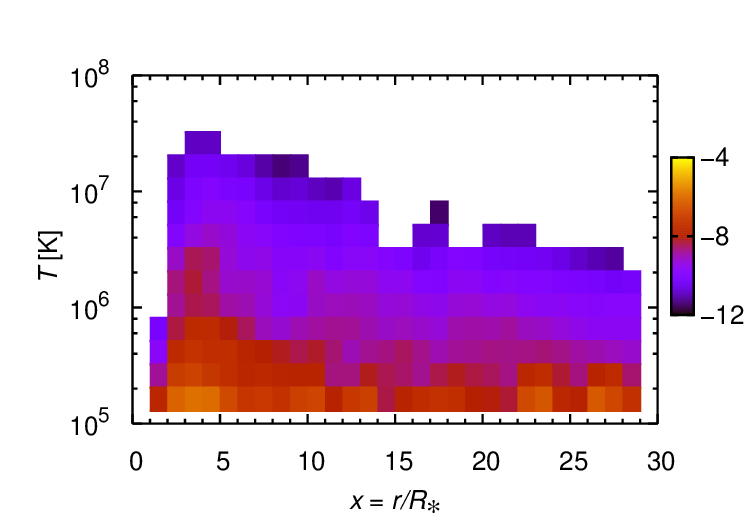}
\includegraphics[width=0.49\hsize]{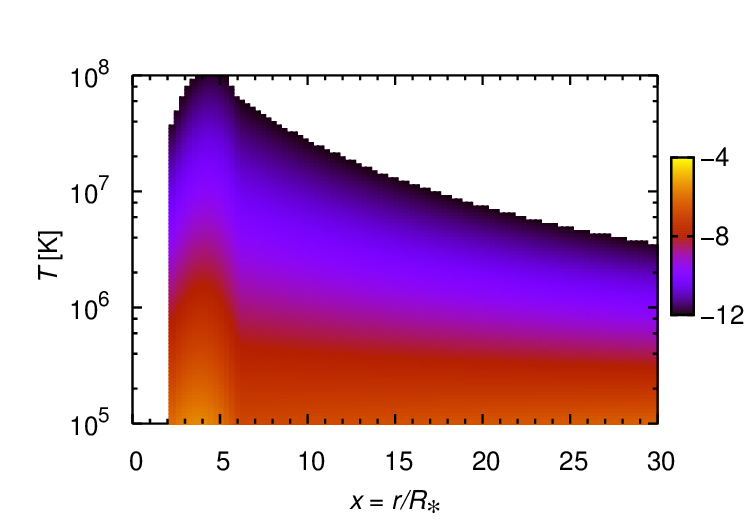}
\caption{{\em Left}: The temperature distribution of X-ray emitting
  gas expressed as $\log\mathcal{D}_{ij}$, as a function of radius.
  {\em Right}: Fit to the temperature distribution expressed as
  $\log\mathcal{D}$.}
\label{hyd2fil_dem}
\end{figure*}

To provide a simpler description of $\mathcal{D}_{ij}$ we approximate
it by a polynomial fit as
\begin{equation}
\label{demfit}
\log\mathcal{D}(x,t)=a(x)+b(x)\,t,
\end{equation}
where $t=\log(T/1\,\text{K})$ is the logarithm of temperature
expressed in units of Kelvin, and $x=r/R_*$. The radial variation of
the coefficients $a(x)$, $b(x)$ is given by
\begin{equation}
\label{decin}
\begin{split}
a(x)&=3.383+1.509x_6-0.279x_6^2+0.416\,(x-x_6),\\
b(x)&=-2.435+0.0564x_6+0.00724x_6^2-0.0783\,(x-x_6).
\end{split}
\end{equation}
Note that $\mathcal{D}=0$ for $x<2$.
For given number densities of free electrons and hydrogen the X-ray
emissivity is
\begin{equation}
\label{etadem}
\eta_\text{X}(r,\nu)= \frac{f^*}{4\pi} n_{\text{e}} n_\text{H}
\int_{T_\text{i}}^{T_\text{f}} \mathcal{D}(r/R_*,\log T)\,\Lambda_\nu(T)\,\de T,
\end{equation}
where we set $T_\text{i}=10^5\,\text{K}$, $T_\text{f}=2.5\times10^7\,\text{K}$,
and $\Lambda_\nu (T)$ is the X-ray emissivity per unit of frequency.
A comparison of $\mathcal{D}$ as derived from the simulations and our
fit is given in Fig.~\ref{hyd2fil_dem}.

We checked that both approximations of X-ray emission from
hydrodynamical simulations (Eq.~\eqref{etax} or \eqref{etadem}) and
models based on exact data give similar results.

\subsection{Scaling of X-ray emissivity with stellar parameters}

The X-ray emission in the hydrodynamical wind simulations of
\cite{felpulpal} was obtained for a mass-loss rate $\mdot_{\zeta\text{
Ori A}}=3\times10^{-6}\,\msr$, radius $R_{\zeta\text{ Ori
A}}=24\,\text{R}_\odot$, and terminal velocity $v_{\zeta\text{ Ori A}}=
1850 \,\kms$ corresponding to a hydrodynamical wind model of
$\zeta$~Ori~A. Most of the X-rays 
used in these simulations
originate
due to the collisions of fast clouds with dense shell fragments, where
the cloud formation is triggered by a turbulent photospheric seed
perturbation. Consequently, it is not clear whether the derived
analytical approximations are realistic also for other stars. To account
for the difference between the X-ray emission in individual star wind
models and $\zeta$~Ori~A model of \citet{felpulpal}, we introduced a
scaling factor $f^*$ in Eqs.~\eqref{etax}, \eqref{etadem}.

The X-ray emissivity is given by the rate of energy dissipation by shocks.
Neglecting possible dependencies on velocity and base perturbations, we expect
the rate of energy dissipation to be proportional to the wind density. However,
this is not what we would get from the scaling $\eta_\text{X}\sim \rho^2$ in
Eqs.~\eqref{etax}, \eqref{etadem} with neglected dependence on the individual
wind parameters, i.e. with $f^*=1$. The reason is that we did not 
take
into
account that the fraction of X-ray emitting material may not be the same for all
stars. In the winds with similar abundances, 
the radiative cooling is more efficient in stars with denser winds than
in stars with low-density winds. In the winds with low 
density, a
larger fraction
of the wind gas may be involved in radiating away the shock thermal energy. 

Based on these 
arguments,
we could introduce the scaling of $f^*$ with
the density $\rho$ at a given point in the wind as $f^*\sim1/\rho$.
However, this is not the most convenient scaling, as it would
introduce additional dependence of $f^*$ on the radius, which is already
accounted for in Eqs.~\eqref{etax}, \eqref{etadem}. Consequently, for each star
the value of $f^*$ should be fixed. Taking into account that
$\rho\sim\mdot/(r^2v)$, we introduce a representative wind density
$\tilde\rho=\mdot/(R_*^2v_\infty)$ and in our calculations use
\begin{equation}
\label{kolin}
f^*= \frac{\tilde\rho_{\zeta\text{ Ori A}}}{\tilde\rho}=
\frac{\mdot_{\zeta\text{ Ori A}}}{\mdot}
\frac{R_*^2}{R_{\zeta\text{ Ori A}}^2}
\frac{v_\infty}{v_{\zeta\text{ Ori A}}},
\end{equation}
where $R_*$, $\mdot$, and $v_\infty$ are radius, wind mass-loss rate,
and terminal velocity of individual stars.

\section{Comparison with the "filling factor" approach}
\label{kapsro}

\begin{figure*}
\centering
\includegraphics[width=0.49\hsize]{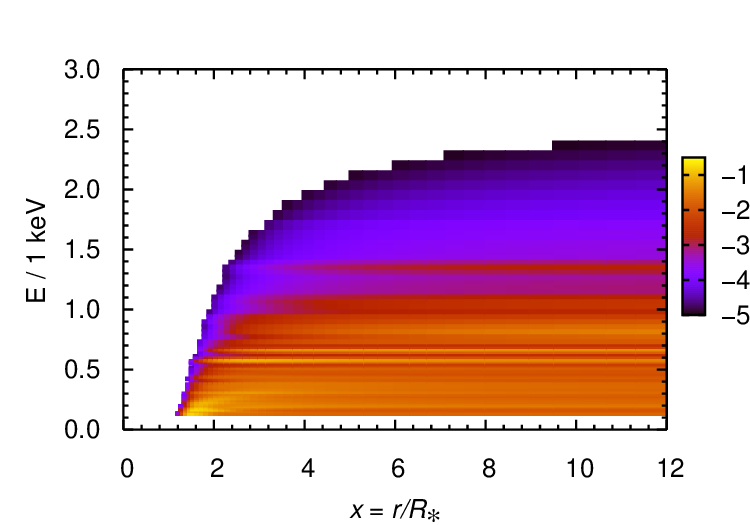}
\includegraphics[width=0.49\hsize]{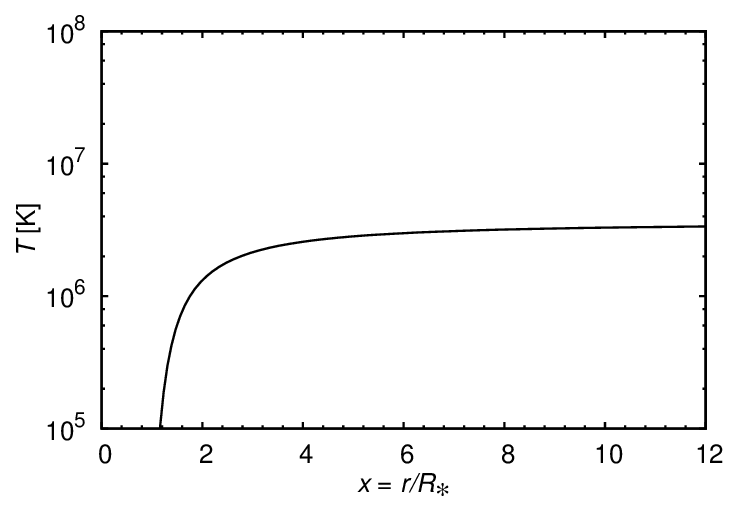}
\caption{{\em Left}: The frequency distribution of X-rays derived
  using the "filling factor" approach expressed in terms of
  $\log (\ell_{ij}/10^{-23}\,\text{erg}\,\text{s}^{-1}\,\text{cm}^{-3}\,
  \text{keV}^{-1})$ as a function of radius and frequency (cf.~with
  Fig.~\ref{hyd2fil}).
  {\em Right}: The temperature of X-ray emitting
  gas expressed as a function of radius 
  (cf.~with Fig.~\ref{hyd2fil_dem}).}
\label{fxkonst}
\end{figure*}

It is informative to compare the X-ray spectrum and the temperature
distribution function derived from hydrodynamic simulations in the
previous section with those derived using simple "filling factor"
approach. The filling factor determines the amount of X-ray emitting
material.

We calculated 
the
wind model of $\zeta$~Ori~A with X-ray emission included after the
"filling factor" approach of \citet{nlteiii} and derived the resulting frequency
distribution of emitted X-rays and the temperature distribution of X-ray
emitting gas (see Fig.~\ref{fxkonst}). The adopted model assumes that the
temperature of the X-ray emitting gas is given by the Rankine-Hugoniot condition
with the velocity difference related to the wind speed. Consequently, the
temperature of X-ray emitting gas in Fig.~\ref{fxkonst} increases with radius
and X-rays are harder in outer wind parts.

This is different from the results of hydrodynamical simulations. In the
simulations the highest temperatures of X-ray emitting gas are achieved
close to the star (see Fig.~\ref{hyd2fil_dem}). As a result, the most
energetic X-rays are emitted close to the star (see Fig.~\ref{hyd2fil}).
Such a
temperature distribution is supported
by 
observations,
as was inferred by \citet{walca} from their extensive
analysis of high-resolution X-ray spectra of O stars. Consequently, two
important features of the X-ray emission emerge from our simulations
(and from observations): the decrease of the temperature of X-ray
emitting gas with radius in 
the outer wind,
and the multitemperature
distribution of X-ray emitting gas at a given point.

\section{Non-LTE models}

\newcommand\mez{\hspace{3mm}}

\begin{table*}
\caption{Stellar parameters of studied O stars.}
\label{obhvezpar}
\centering
\begin{tabular}{rrcccccc}
\hline
\multicolumn{1}{c}{Star} & \multicolumn{1}{c}{HD} & Sp. & ${R_{*}}$ & $M$ &
$\Teff$ & $L$ & $\dot M$\\
& \multicolumn{1}{c}{}& type & $[\text{R}_{\odot}]$ &$[\text{M}_{\odot}]$
&  $[10^3\,\text{K}] $ & $[10^5\,\text{L}_\odot]$ & $[\msr]$ \\
\hline
$\xi$ Per      & $24912$&O7.5IIIe& $14.0$ & $36$ & $35.0$ & $2.64$ & $4.6\times10^{-7}$\\
$\alpha$ Cam   & $30614$&O9.5Iae & $27.6$ & $43$ & $30.9$ & $6.23$ & $1.5\times10^{-6}$\\
$\lambda$ Ori A& $36861$&O8 III  & $12.3$ & $30$ & $36.0$ & $2.28$ & $5.1\times10^{-7}$\\
$\iota$ Ori A  & $37043$&O9III   & $21.6$ & $41$ & $31.4$ & $4.07$ & $6.3\times10^{-7}$\\
$\zeta$ Ori A  & $37742$&O9Iab   & $24.0$ & $34$ & $31.5$ & $5.09$ & $1.5\times10^{-6}$\\
15 Mon         & $47839$&O7Ve    &  $9.9$ & $32$ & $37.5$ & $1.74$ & $2.8\times10^{-7}$\\
               & $54662$&O7III   & $11.9$ & $38$ & $38.6$ & $2.82$ & $8.9\times10^{-7}$\\
               & $93204$&O5V     & $11.9$ & $41$ & $40.0$ & $3.25$ & $1.4\times10^{-6}$\\
$\zeta$ Oph    &$149757$&O9V     &  $8.9$ & $21$ & $32.0$ & $0.75$ & $4.6\times10^{-8}$\\
63 Oph         &$162978$&O8III   & $16.0$ & $40$ & $37.1$ & $4.35$ & $2.0\times10^{-6}$\\
68 Cyg         &$203064$&O8e     & $15.7$ & $38$ & $34.5$ & $3.13$ & $6.1\times10^{-7}$\\
19 Cep         &$209975$&O9Ib    & $22.9$ & $47$ & $32.0$ & $4.93$ & $8.6\times10^{-7}$\\
$\lambda$  Cep &$210839$&O6Iab   & $19.6$ & $51$ & $38.2$ & $7.34$ & $6.4\times10^{-6}$\\
\hline AE Aur  &$34078$ &O9.5Ve  & $ 7.5$ & $20$ & $33.0$ & $0.60$ & $1.4\times10^{-8}$\\
$\mu$ Col      &$38666$ &O9.5V   & $ 6.6$ & $19$ & $33.0$ & $0.46$ & $7.8\times10^{-9}$\\
               &$42088$ &O6.5V   & $ 9.6$ & $31$ & $38.0$ & $1.72$ & $3.4\times10^{-7}$\\
               &$46202$ &O9V     & $ 8.4$ & $21$ & $33.0$ & $0.75$ & $2.3\times10^{-8}$\\
\hline
\end{tabular}
\end{table*}

\subsection{Model assumptions}
\label{modas}

As an application of derived analytical formulae, we include X-ray
emission parameterised by Eqs.~\eqref{etadem} and \eqref{kolin} into our
stationary, spherically symmetric non-LTE wind model \citep[see][for a
detailed description]{nlteiii}. The X-ray emission is included in the
radiative transfer equation of our models. In the following we discuss
just the temperature fitting approach as it seems to be more
straightforward. The hydrogen and free electron densities in the X-ray
source term are taken from the non-LTE models.

The excitation and ionization state of the considered elements is
calculated from statistical equilibrium (non-LTE) equations. We use
atomic data from the Opacity Project and the Iron Project (Seaton
\citeyear{top}, \citeauthor{opc5} \citeyear{opc5}, Luo \& Pradhan
\citeyear{top1}, Sawey \& Berrington \citeyear{savej}, Seaton et
al. \citeyear{topt}, Butler et al. \citeyear{bumez}, Nahar \& Pradhan
\citeyear{napra}, Hummer et al. \citeyear{zel0}, Bautista
\citeyear{zel6}, Nahar \& Pradhan \citeyear{zel2}, Zhang
\citeyear{zel1}, Bautista \& Pradhan \citeyear{zel5}, Zhang \& Pradhan
\citeyear{zel4}, Chen \& Pradhan \citeyear{zel3}). A significant part
of the ionic models was taken from the TLUSTY code
\citep{ostar2003,bstar2006}. For phosphorus we employed data described
by Pauldrach et al. (\citeyear{pahole}).
Auger photoionization cross sections from individual inner-shells were taken
from  \citet[see also \citealt{muzustahnouthura}]{veryak}, and Auger yields were
taken from \citet{kame}. We use \citet{asgres} solar
abundance determinations in the code.

The radiative transfer equation is split 
in
two parts, i.e., the
continuum and line radiative transfer. The radiative transfer in the
continuum is solved using the Feautrier method in spherical
coordinates (Mihalas \& Hummer \citeyear{sphermod} or Kub\'at
\citeyear{dis}) with inclusion of all free-free and bound-free
transitions of the model ions. The radiative transfer in lines is
solved in the Sobolev approximation (e.g., Castor \citeyear{cassob})
neglecting continuum opacity and line overlaps.

The radiative force is calculated in the Sobolev approximation (see
Castor \citeyear{cassob}). The corresponding line data were extracted
in 2002 from the VALD database (Piskunov et al. \citeyear{vald1},
Kupka et al. \citeyear{vald2}). The radiative cooling and heating
terms are calculated using the electron thermal balance method
(Kub\'at et al., \citeyear{kpp}). For the calculation of these terms
we use occupation numbers derived from the statistical equilibrium
equations. Finally, the continuity equation, equation of motion, and
energy equation are solved iteratively to obtain the wind density,
velocity, and temperature structure.

The shortest wavelength considered in the models is $4.1$\,\AA\ and the
longest X-ray wavelength is defined 
as
$100$\,\AA.

\subsection{Test stars}

The purpose of our study is to investigate the trends in the X-ray
emission known for all OB stars, rather than to study X-ray properties of
individual objects. Therefore, we compiled a list of O stars that includes
the objects of various luminosity classes, masses, and binarity status. By
doing this, our list may be considered as 
%
representative of
a
diverse
population of O stars found in real star clusters. This allows
us
to
compare the trend in X-ray luminosity of our synthetic O star population
with the real clusters recently observed by \citet{sane} and \citet{igorkar}.

To address the ``weak wind problem'' \citep{bourak,martclump}, we also
included stars that should be subject to it, i.e., their observed wind
lines are much weaker than expected (the bottom four stars below the
horizontal line in Table~\ref{obhvezpar}).

The stellar parameters (see Table \ref{obhvezpar}) are taken from
Lamers et al. (\citeyear{lsl}), Repolust et al. (\citeyear{rep}),
Markova et al.  (\citeyear{upice}), and \citet{martclump}. Note that
the parameters derived by Repolust et al. (\citeyear{rep}), Markova et
al. (\citeyear{upice}), and \citet{martclump} were obtained using
blanketed model atmospheres, i.e., they should be more reliable than
the older ones. Stellar masses were obtained using evolutionary tracks
(Schaller et al.~\citeyear{salek}). The mass-loss rates in
Table~\ref{obhvezpar} were theoretically predicted
using our models discussed in Sect.~\ref{modas}.

\section{Non-LTE models with X-ray emission from simulations}
\label{kapnltemod}

The inclusion of X-ray emission does not significantly modify the
ionization fractions close to the star below the critical point radius
where the mass-loss rate is fixed. Consequently, X-rays do not
significantly modify the wind mass-loss rate, but may modify the wind
terminal velocity by a few percent \citep{nlteiii}.

\subsection{X-ray luminosity}

\begin{figure}
\centering
\includegraphics[width=0.9\hsize]{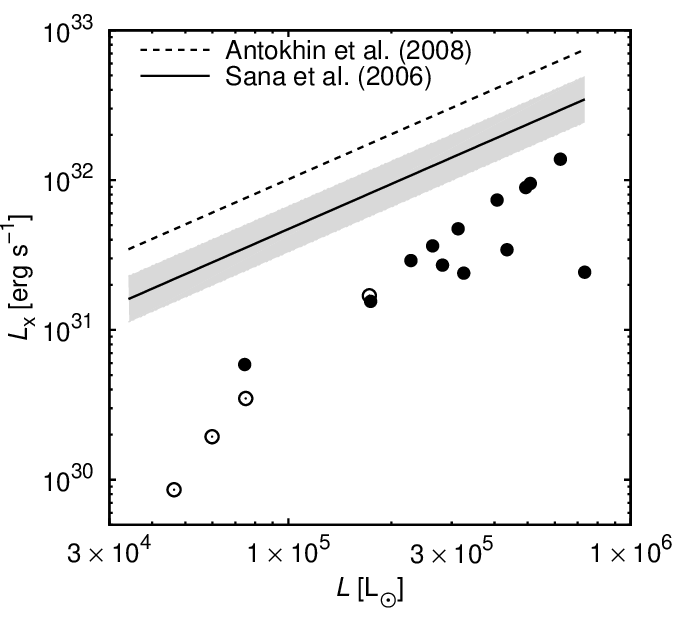}
\caption{The dependence of the total X-ray luminosity on the
  bolometric luminosity calculated using non-LTE models with X-ray
  emissivity after Eq.~\eqref{etadem} and Eq.~\eqref{kolin} 
  (filled and empty circles) for individual stars. This is
  compared with the mean observational relations
  (corrected for the interstellar absorption) derived by
  \citet{igorkar} and \protect\citet[with uncertainty]{sane}.
  Empty circles denote values for stars showing the "weak wind problem".}
\label{lxlbolhv}
\end{figure}

A comparison of the predicted X-ray luminosities for individual stars
and the mean observational trends is given in Fig.~\ref{lxlbolhv}. The
predicted X-ray luminosities for stars with optically thick winds
($L\gtrsim10^5\,\text{L}_\odot$) are on average
lower
roughly by a factor of
three 
%
than the observed ones. This corresponds to the results of
\citet{felpulpal}.

The derived slope of the $\lx-L$ relation for stars with optically thick
winds, $\lx\sim L^{1.0}$, is in 
%
better agreement with observations.
The cause of the decrease of the slope (compared 
to
results obtained
for a fixed filling factor as a free parameter, \citealt{nlteiii}) can
be most easily understood from Fig.~\ref{hyd2fil}, which shows that
$\ell(x,\gamma)$ decreases with $x$ (i.e., with radius). Consequently,
the filling factor also decreases with radius, supporting the
explanation of \citet{oskal}. Also the dependence of $f^*$ on the
stellar parameters contributes to the decrease of the slope of the
$\lx-L$ relation, but the contribution of this parameter is sensitive to
the stellar sample considered.

For stars with optically thin wind 
($L\lesssim10^5\,\text{L}_\odot$),
even the improved inclusion of X-ray emission is not capable of
explaining the observed slope of the $\lx-L$ relation. As indicated
above, 
the dependence of
the predicted X-ray luminosity of these stars 
on 
%
stellar luminosity 
is
steeper than for stars with high luminosity. However,
this is not supported by observations. The $\lx-L$ relation of low
luminosity stars is either the same as that of high luminosity stars,
or, for B stars, the slope is even less steep \citep{sane}. The reason
for this discrepancy between observation and theory is unclear. Note
that for these stars the shock cooling length could be comparable with
the hydrodynamical scale \citep{cobecru,nlteiii}.  The fact that a
substantial fraction of the wind is too hot to give a significant
signature in the ultraviolet spectrum of the star could be
an explanation for
the weak-wind problem \citep{nlteiii}.

\subsection{The energy distribution of X-rays}

\begin{figure}
\centering
\resizebox{0.8\hsize}{!}{\includegraphics{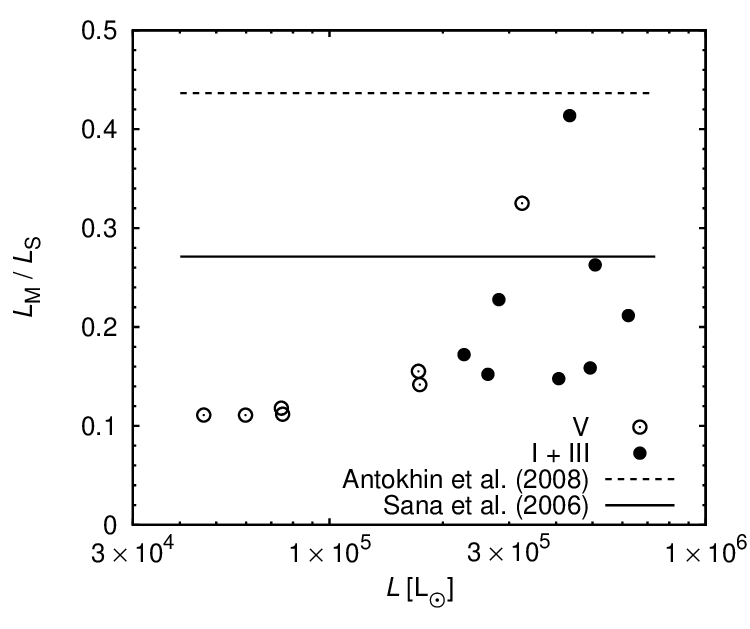}}
\caption{Comparison between the predicted ratio of X-ray
  luminosities in the medium and soft energy bands
  for individual stars and the mean trend derived from observations.
  Main sequence
  stars are denoted with empty symbols, giants and supergiants with
  filled symbols.}
\label{lxlbolhv_mek}
\end{figure}

\citet{sane} and \citet{igorkar} divide the X-rays into three energy
bands, soft ($0.5-1.0\,$keV), medium ($1.0-2.5\,$keV), and hard
($2.5-10.0\,$keV). The observations show that the slope of the $\lx-L$
relation is roughly the same in the soft and medium band as in the
total X-ray emission, whereas the observations in the hard band show a
large dispersion.

Our calculations show a somewhat different result, as can be seen from
Fig.~\ref{lxlbolhv_mek}, where we plot the ratio of the X-ray
luminosities in the medium ($L_\text{M}$) and soft ($L_\text{S}$)
energy bands. The predicted slope of the $\lx-L$ relation in the
medium band is steeper than that in the soft band, consequently the
ratio of $L_\text{M}/L_\text{S}$ increases with luminosity. The reason
for this behaviour is the inverse proportionality of X-ray opacity and
energy \citep[e.g.][]{xlida,nlteiii}. For stars with low 
luminosities,
the wind density is 
low,
and most of the emitted X-rays will be
emergent. On the other hand, for stars with 
high
luminosities,
density and opacity
values
become large, more flux is absorbed in the soft
than in the medium energy band, consequently the emergent X-rays become
harder. This also explains why the predicted hardness of X-rays is
greater
for giants and supergiants than for main sequence stars (see
Fig.~\ref{lxlbolhv_mek}). Finally, although from our models 
it
follows that
the predicted dispersion of
X-ray luminosities is indeed 
higher
in the hard band than in the
medium and soft energy bands, there is still a clearly discernible
$\lx-L$ relation even in the hard band.

\subsection{Ionization fractions}

\begin{figure*}
\centering
\resizebox{0.45\hsize}{!}{\includegraphics{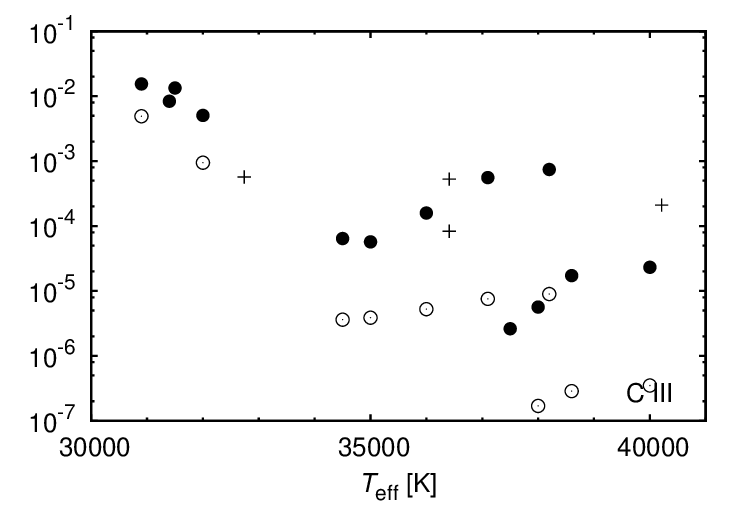}}
\resizebox{0.45\hsize}{!}{\includegraphics{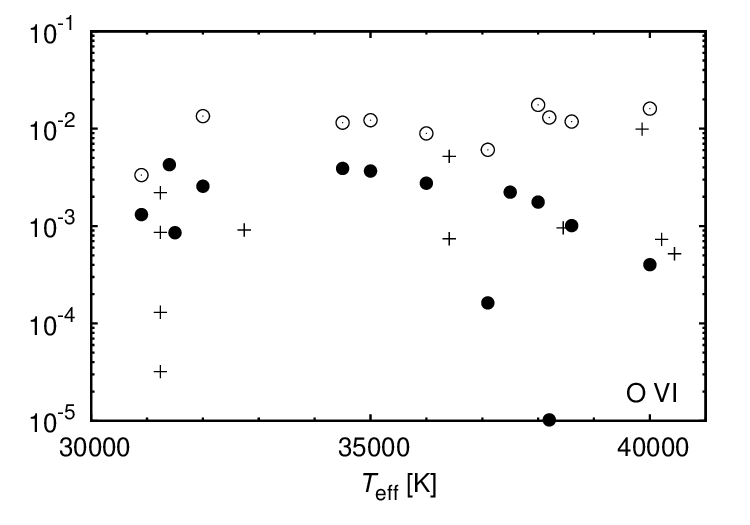}}
\caption{Ionization fractions as a function of the effective
  temperature for individual stars from our sample (only stars with
  $\dot M>10^{-7}\,\msr$ are included here) at the point where the
  radial velocity $v=0.5v_\infty$. Filled circles
  {\lower.3ex\hbox{\LARGE\textbullet}} refer to the present models
  with X-ray emission from hydrodynamical simulations, open circles
  $\odot$ denote values taken from former non-LTE models with X-ray
  emission described using filling factor \citep{nlteiii}. The
  ionization fractions derived from observations were adopted from
  \protect\citet[for LMC stars, plus signs $+$]{maso}.}
\label{iontep}
\end{figure*}

The presence of X-rays may also influence the ionization fractions of
highly ionised ions. We plot in Fig.~\ref{iontep} the predicted
ionization fractions in comparison with those derived from observation
as a function of effective stellar temperature. Note that 
here
we use the
ionization fractions derived for LMC stars \citep{maso},
%
because
their observational sample includes a large number of ionization stages,
the fractions
of which are given as a function of
%
velocity, and the wind parameters of LMC stars are not
significantly different from 
%
Galactic ones. Since several parameters
except the effective temperature (e.g., the wind density) may influence
the ionization fractions, the graphs in Fig.~\ref{iontep} are not
monotonic.

In Fig.~\ref{iontep} we also plot the results of \citet{nlteiii}, which
were obtained using a simpler inclusion of X-rays assuming a constant
filling factor (see also Sect.~\ref{kapsro}). Still, the emergent X-ray
luminosities from these models roughly correspond to the observed ones.
Rather surprisingly, although the non-LTE models calculated with X-ray
emission from hydrodynamical wind simulations give a too low emergent
X-ray luminosity, the ionization structure of these models corresponds
in general much better to the trends derived from 
observations.

This better agreement between theoretical and observational ionization
fractions may be due to an overall lower X-ray emissivity. To test
this, we calculated models with a simplified treatment of X-ray
emission after \citet{nlteiii}, but assuming the same X-ray luminosity
as that predicted by hydrodynamical simulations. Our results show that
even in this case the use of X-ray emissivity based on hydrodynamical
simulation gives better agreement with observations. We
conclude that the hydrodynamical simulations give a more realistic
dependence of the X-ray emissivity on frequency.

\section{Discussion}

The remaining discrepancy between theory and observation may originate
in our neglect of macroclumping (sometimes 
also termed porosity).
The
macroclumping may cause a lower opacity in the X-ray domain, which leads
to a 
higher
predicted X-ray luminosity. From Fig.~(17) in \citet{osfeha},
it
follows that 
an
increase of X-ray luminosity by a factor of 3 would be
easily achievable. Accounting for macroclumping reduces the wavelength
dependence of opacity.  In its limit (when clumps are fully opaque), the
opacity becomes grey \citep{osfeha}. 

Wind inhomogeneities should also affect the ionization fractions.
Optically thin inhomogeneities (often referred to as ``microclumping'') lead
generally to lower ionization stages \citep{lijana,mychuch}, hence
clumping 
%
with increased X-ray emissivity could be another way
to explain the X-ray luminosities and ionization fractions derived
from observation.

Our
hydrodynamical wind simulations assume
1-D spherical symmetry, thus all wind structures correspond to full
spherical shells. Therefore, we have at present a very limited
knowledge of real wind clumping.

To be consistent with the hydrodynamical simulations, we use 
%
the
Raymond-Smith X-ray spectral code (Raymond \& Smith \citeyear{rs},
Raymond \citeyear{ray}), although it would be possible to use codes
based on up-to-date atomic data. Our test using the APEC subroutine
available in XSPEC \citep{xspec,apec} showed that there is a relatively
good agreement between the fits
%
using this
and the Raymond-Smith code
Eq.~\eqref{zosen}.

\section{Summary}

We provide analytical approximations 
of
the X-ray emission predicted by
hydrodynamical simulations of hot star winds by \citet{felpulpal}. The
X-ray emission derived from the hydrodynamical simulations has two
distinctive aspects. First, the temperature of X-ray emitting gas
decreases with radius in 
the
outer wind.
%
Second, the temperature of
the X-ray emitting gas is described by a distribution function, which is
more realistic than assuming just one temperature at a given point.

We include these approximations 
in
non-LTE models \citep{nlteiii} and
for selected stars compare the resulting X-ray luminosities, energy
distribution of emergent X-rays, and ionization structure with
observational results. We conclude that although the predicted X-ray
luminosity is slightly underestimated, the resulting ionization
structure is in relatively good agreement with observations. Earlier
papers debated whether the theory of line-driven winds can explain the
observed 
$\lx\sim L$ relation.
Our results reproduce this scaling.

\begin{acknowledgements}
We thank the referee for his/her comments that helped to improve the manuscript.
This work was supported by grant GA \v CR 205/08/0003. LMO acknowledges
the support of DLR grant 50\,OR\,0804.
\end{acknowledgements}

\end{document}